\documentclass[11pt,a4paper]{article}
\usepackage[margin=1.1in]{geometry}
\pdfoutput=1

\usepackage{epsfig}
\usepackage{latexsym}
\usepackage{enumerate}
\usepackage{url}
\usepackage{float}
\usepackage[hypertexnames=false,hyperfootnotes=false]{hyperref}
\usepackage[T1]{fontenc}
\usepackage{color}
\usepackage{tikz}
\usepackage[margin=10pt,font=small,labelfont=bf]{caption}
\usepackage[subrefformat=parens,labelformat=parens]{subcaption}
\usepackage{afterpage}
\usepackage{enumitem}

\usepackage{lmodern}
\usepackage{booktabs}
\usepackage{sectsty}
\usepackage{ifthen}
\usepackage{amsmath}
\usepackage{amsthm}
\usepackage{amssymb}
\usepackage{amsfonts}
\usepackage{thmtools}
\usepackage{thm-restate}
\usepackage{mathtools}
\usepackage{xspace}
\usepackage{titling}
\usepackage[numbers,sort&compress]{natbib}
\usepackage{xfrac}
\usepackage{multirow}
\usepackage{bigdelim}
\usepackage{bm}
\usepackage{cleveref}
\usepackage{microtype}
\usepackage{xurl}
\hypersetup{hidelinks}
\newcommand{\field}[1]{\ensuremath{\mathbb{#1}}}
\newcommand{\R}{\ensuremath{\field{R}}} % real numbers
\newcommand{\I}[1]{\ensuremath{\mathbb{I}_{\left\{#1\right\}}}} % indicator function
\newcommand{\PR}{\ensuremath{\mathsf{P}}} % probability
\newcommand{\E}{\ensuremath{\mathsf{E}}} % expectation

\newcommand{\Fscr}{\ensuremath{\mathcal F}}

\newcommand{\Oscr}{\ensuremath{\mathcal O}}

\newcommand{\Xscr}{\ensuremath{\mathcal X}}

\DeclarePairedDelimiter\norm{\lVert}{\rVert}

\DeclareMathOperator{\Var}{Var}
\DeclareMathOperator{\Cov}{Cov}

\DeclareMathOperator*{\argmin}{\mathrm{argmin}}

\declaretheoremstyle[headfont=\sffamily\bfseries,bodyfont=\itshape]{thm-sf}
\declaretheorem[style=thm-sf]{theorem}
\declaretheorem[style=thm-sf]{remark}

\crefname{assumption}{assumption}{assumptions}

\declaretheorem[style=thm-sf]{example}

\declaretheorem[style=thm-sf]{lemma}

\declaretheorem[style=thm-sf]{proposition}

\renewcommand{\thmcontinues}[1]{\hyperref[#1]{continued}}

\usetikzlibrary{arrows,patterns,plotmarks,pgfplots.groupplots}
\tikzstyle{every picture} += [>=stealth]
\tikzset{axis/.style={semithick, line join=miter}}
\allsectionsfont{\sffamily}
\makeatletter
\def\@seccntformat#1{\csname the#1\endcsname.\quad}
\makeatother

\floatstyle{ruled}
\provideboolean{fastcompile}
\newcommand{\hidefastcompile}[1]{\ifthenelse{\boolean{fastcompile}}{}{#1}}
\usepackage{pgfplots}
\usepackage{pgfplotstable}
\usetikzlibrary{calc}
\usepackage{mathtools}
\definecolor{orange}{rgb}{0.85,0.33,0.13} % 217,85,33
\definecolor{green}{rgb}{0.13,0.85,0.33}
\definecolor{purple}{rgb}{0.33,0.13,0.85}
\definecolor{lime}{rgb}{0.65,0.85,0.13}
\definecolor{blue}{rgb}{0.13,0.65,0.85}
\pgfplotscreateplotcyclelist{tricolor}{%
  orange,every mark/.append style={fill=orange!80!black},mark=*\\%
  green,every mark/.append style={fill=green!80!black},mark=square*\\%
  purple,every mark/.append style={fill=purple!80!black},mark=otimes*\\%
  black,mark=star\\%
  orange,every mark/.append style={fill=orange!80!black},mark=diamond*\\%
  green,densely dashed,every mark/.append style={solid,fill=green!80!black},mark=*\\%
  purple,densely dashed,every mark/.append style={solid,fill=purple!80!black},mark=square*\\%
  black,densely dashed,every mark/.append style={solid,fill=gray},mark=otimes*\\%
  orange,densely dashed,mark=star,every mark/.append style=solid\\%
  green,densely dashed,every mark/.append style={solid,fill=green!80!black},mark=diamond*\\%
}
\pgfplotsset{colormap={tricolormap}{color=(orange) color=(green) color=(purple)},
  colormap={quadcolormap}{color=(orange) color=(lime) color=(blue) color=(purple)}}
\pgfplotstableset{%
  font=\small,
  every head row/.style={before row=\toprule[1pt], after row=\midrule},
  every last row/.style={after row=\bottomrule[1pt]}}
\pgfplotsset{compat=1.15}

\DeclareMathOperator{\MMSE}{MMSE}

\newcommand{\draftnewpage}{}
\newcommand{\draftpagebreak}{}

\title{The Geography of (Concurrent) Block Proposing}
\author{%
  Ciamac C. Moallemi\thanks{Columbia University and Paradigm.}\quad
  Mallesh M. Pai\thanks{Rice University and Tempo Labs.}\\[0.5em]
  Dan Robinson\thanks{Paradigm}\quad
  Christoph Schlegel\thanks{Flashbots}%
}
\date{October 2026}

\begin{document}

\maketitle
\begin{abstract}
The resilience and
censorship resistance of a blockchain depend on where the validators and proposers of that blockchain physically operate. We study the economic incentives behind this location choice, both with a single
proposer and with multiple concurrent proposers. Formally, we study a game where proposers first
choose locations and then compete in a first-price auction for an opportunity whose common value is
driven by information from geographically dispersed sources. A single proposer minimizes
variance-weighted delay, locating at a weighted median of the sources and co-locating with any
dominant one. By contrast, with two proposers, information both observe equally early earns no
rent, and each proposer's expected profit is proportional to the standard deviation of the
information it receives before its rival. We show that the location game admits the jointly
unobserved variance as an ordinal potential, so global solutions of a weighted two-median problem
are equilibria.
With two sources, one proposer locates at each, however weak one is. With more proposers and at
least as many sources, equilibrium locations are distinct under regular bidding. Multiple
concurrent proposers thus do more than limit any single proposer's control over inclusion: by
rewarding exclusive rather than shared information, they create an economic force toward
geographic decentralization.
\end{abstract}

\draftnewpage

\section{Introduction}
\label{sec:introduction}

Geographic decentralization matters for the resilience of a blockchain. A blockchain can be
operated by a collective of independent agents, but if those agents operate from the same
location, they share that location's electricity supply, communications infrastructure, and
exposure to natural disasters. A local disruption can remove many of them at once. Concentration
within one jurisdiction likewise exposes otherwise independent agents to the same political or
regulatory pressure to exclude transactions or cease operating. Geographic dispersion reduces
these shared vulnerabilities and preserves alternative routes for transaction inclusion when some
agents become unavailable or censor. It is a distinct dimension of decentralization from the
distribution of ownership or stake~\cite{gencer2018,yang2026}: adding agents offers little
protection against a regional shock if they all locate in the same region.

There is also a distributional case for geographic decentralization. If profitable participation
requires proximity to a small set of hubs, participants elsewhere face higher communication costs
or must relocate their infrastructure. A broader geographic footprint can spread access to
low-latency services and opportunities to earn revenue. 

Geography matters in financial markets, because information is produced at particular places
and reaches other places with delay. A participant located near a market where information
originates learns of a trading opportunity before participants located elsewhere. The economic
value of reducing this delay is well established in trading~\cite{moallemi2013latency,budish2015}.
For example a timing advantage in a common-value auction generates rents for the faster
bidder~\cite{moallemi2026latency}. Blockchain proposers face the same considerations. A proposer
learns about an opportunity sooner by operating near the market where the relevant information
originates, and it may also benefit from proximity to the validators or intermediaries that must
receive its messages. Location therefore has an economic role even when the protocol does not
consider geography explicitly.

Geography is not set by the protocol designer, however. Participants choose where to operate in
response to economic incentives, and the resulting geography is an equilibrium outcome. If earning
proposer revenue requires proximity to a particular market or network hub, nominally open
participation can coexist with strong pressure to operate in a narrow set of locations, and
increasing the number of participants need not spread them out if all of them benefit from moving
toward the same sources. This question is particularly relevant for multiple concurrent proposer
designs, which allow several proposers to contribute within the same block-production interval in
order to limit the control any single proposer has over transaction inclusion and
ordering~\cite{fox2023censorship,garimidi2025mcp,cadence}. The geographic consequences of these designs
depend on the incentives of the participants. We study the location incentives created when concurrent proposers compete for the same
financial opportunity. The proposer is operated together with a searcher, or its return is aligned
with that searcher's profit. The question is whether competition induces these participants to
gather around the most valuable information source or to establish informational advantages in
different places. We answer by making their locations endogenous and deriving the information
rents that support them.

Our central mechanism is the change in the value of information when a proposer faces a
competitor. A monopolist benefits from improving its estimate of an opportunity's value. With
competing proposers, information that reaches both participants equally early is common to them
and creates no informational advantage. A proposer instead earns rents from the information it
receives ahead of its rival. Moving toward a source that is less important in aggregate can
therefore be profitable if doing so creates an exclusive informational lead. The geography of
competition depends on the rival's location as well as the distribution of information sources.

Proposers can earn revenue from financial opportunities whose value depends on external prices, such as filling a user’s order or trading first against an automated market maker. When concurrent proposers compete for the same opportunity, their profits depend on what they know about its value relative to their rivals. Geography matters because price information from nearby trading venues arrives sooner.

We model this competition as a two-stage game. Proposers first choose locations and then compete in a first-price auction for a right with a common terminal value. This value evolves through independent Brownian innovations originating at geographically dispersed sources. Propagation delays determine which innovations each proposer observes before bidding. Section~\ref{sec:interpretation} explains the auction’s interpretation and how searcher profits translate into proposer location incentives.

\paragraph{Contributions.} Our contributions are as follows:
\begin{itemize}
\item \emph{Rents come from exclusive information.} The auction has a closed-form equilibrium in
  which each proposer's expected profit is proportional to the standard deviation of the
  information it observes before its rival (\Cref{thm:quantile} and \Cref{cor:gaussian}).

\item \emph{A monopolist concentrates; competitors separate.} A single proposer locates at a
  variance-weighted median of the sources (\Cref{thm:monopoly}). With two proposers, the jointly
  unobserved variance is an ordinal potential; equilibria exist, and global solutions of the
  associated weighted two-median problem are equilibria
  (\Cref{thm:potential}). Co-location requires one location that minimizes every source's
  delay (\Cref{cor:colocation}); with two sources, one proposer locates at each regardless of
  their weights (\Cref{prop:two-sources}).

\item \emph{More proposers.} With at least as many sources as proposers, equilibria with regular
  bidding have pairwise distinct locations, though possibly close together (\Cref{thm:many-distinct}); existence of such equilibria is open.

\item \emph{Protocol design matters.} Location-dependent deadlines enter as effective delays
  (\Cref{prop:effective-delay}), and a single fixed auctioneer makes co-location an equilibrium
  (\Cref{prop:auctioneer}).

\item \emph{Price informativeness.} Competition makes the auction outcome more informative about
  the value than an optimally located monopolist (\Cref{thm:information-comparison}), but
  equilibrium locations need not be the most informative ones
  (\Cref{prop:information-locations}).
\end{itemize}

\subsection{Related literature}
\label{sec:related-literature}

\paragraph{Geographic decentralization.} Gencer et al.~\cite{gencer2018} measure how network
resources and interconnection shape decentralization in Bitcoin and Ethereum, a dimension distinct
from the distribution of ownership or stake. Closest to our question, Yang et
al.~\cite{yang2026} link validator positioning to information sources, propagation, and the
choice between local and external block building, using latency-calibrated simulations. We complement this reduced form model with an explicit model of the information available to proposers and the auction through which they compete, deriving location incentives from the resulting equilibrium profits.

Babel, Milionis, and Xue~\cite{babel2026geographic} also study geographic incentives under
multiple concurrent proposers. Their proposers maximize transaction market share: users route
transactions according to inclusion time, which includes delivery to an attester quorum.
We instead derive location payoffs from auction rents on geographically dispersed information.

Roeschlin et al.~\cite{roeschlin2026diversity} design explicit rewards for geographic diversity,
supported by location verification and incentives for truthful reporting. Milionis et
al.~\cite{milionis2025recovery} study truthful recovery of information from manipulable signals,
including location. Our model takes locations and delays as public and studies where participants
choose to operate without geography-based rewards.

\paragraph{Multiple concurrent proposers.} Fox, Pai, and Resnick~\cite{fox2023censorship} show
that a single proposer can censor competing bids in on-chain auctions, and use conditional tips
to create competition among concurrent proposers. Garimidi, Neu, and
Resnick~\cite{garimidi2025mcp} and Cadence~\cite{cadence} implement such competition. These
papers motivate and implement proposer competition; we study the geographic incentives it creates, and abstract from consensus
guarantees and implementation. Schwarz-Schilling et al.~\cite{timinggames} study a related
trade-off in time, between waiting for additional value and delivering a proposal on time. In
our model, location plays the role of timing: it determines which information arrives before the
deadline.

\paragraph{Latency in markets.} The value of speed in trading is well established. Moallemi and
Sa\u{g}lam~\cite{moallemi2013latency} quantify the cost of delayed market access in an optimal
execution problem and relate it to volatility and the bid--ask spread. Budish, Cramton, and
Shim~\cite{budish2015} show that continuous trading induces an arms race in speed and propose
frequent batch auctions in response. Schlegel~\cite{schlegel2026ordering} studies bidding and
endogenous latency investment under batch auctions and time-boost ordering, showing that batch
auctions can retain incentives to invest in speed. As in those papers, the private value of speed here depends
on the market rules and on how information evolves. Our choice variable differs: a location
relative to several sources and to competitors, with the propagation technology fixed, rather
than an investment in faster infrastructure.

\paragraph{Common-value auctions with asymmetric information.} Milgrom and
Weber~\cite{milgromweber} provide the general framework for auctions with interdependent values. Engelbrecht-Wiggans, Milgrom, and
Weber~\cite{engelbrechtwiggans1983} characterize bidding when one bidder is privately informed
and its rivals hold only public information. Moallemi, Pai, and Robinson~\cite{moallemi2026latency}
characterize the equilibrium with a reserve when the informed bidder's advantage comes from
bidding later, and quantify the value of that timing advantage. Our one-sided case
(\Cref{prop:degenerate}) is that of Engelbrecht-Wiggans et al.~\cite{engelbrechtwiggans1983}, and
the bidding formula in \Cref{thm:quantile} uses the quantile construction of Cheng and
Tan~\cite{chengtan2010} for asymmetric first-price common-value auctions. Our model makes
these information differences endogenous to geographic location: each proposer can be ahead
at a different source.

\paragraph{Potential games and facility location.} We use the exact and ordinal potentials of
Monderer and Shapley~\cite{monderershapley}. A monopolist solves a weighted-median location
problem; under competition, because each proposer's rent is a monotone transform of its exclusive
variance, the residual variance becomes an ordinal potential and the location problem becomes a
two-median coverage problem. Global two-median solutions are equilibria, but the equilibrium set
can also contain coordinatewise minima that are not global.

\section{Model}
\label{sec:model}

\subsection{Sources, locations, and information}

There are $M\geq 1$ proposers and $K\geq 1$ information sources. Proposer $i$ chooses a location
$x_i$ in a compact metric space $\Xscr$. Source $k$ is at $z_k$. Its information reaches location
$x$ with delay $d_k(x)$, where $d_k:\Xscr\to[0,T]$ is continuous and $T>0$ is the settlement
time. In the distance examples, source sites belong to $\Xscr$ and $d_k(x)=\rho(z_k,x)$ for a
metric $\rho$, expressed in units of time, with $\rho(z_k,x)\leq T$ for all $k$ and $x\in\Xscr$. Section~\ref{sec:deadlines} allows the delay to include
a location-dependent bidding deadline.

The terminal value is common to all proposers and given by
\begin{equation}
 V_T=V_0+\sum_{k=1}^K\sigma_k B^k_T,
 \qquad w_k:=\sigma_k^2>0,\qquad W:=\sum_{k=1}^K w_k,
 \label{eq:value}
\end{equation}
where $V_0\in\R$ is public and the $B^k$ are independent standard Brownian motions. A proposer observes the history received from each source, but with delay. At location $x_i$ its information at time $T$ is
\begin{equation}
 \Fscr_i(x_i):=\sigma\bigl(B^k_t:0\leq t\leq T-d_k(x_i),\ 1\leq k\leq K\bigr).
 \label{eq:information}
\end{equation}
Locations, delays, and source variances are public. A faster proposer can recover the older observation available to the slower proposer.

For a single location define its \emph{variance-weighted delay} and \emph{observed variance} by
\begin{equation}
 D(x):=\sum_k w_kd_k(x),\qquad I(x):=TW-D(x).
 \label{eq:delay-cost}
\end{equation}
The former is the variance still unobserved at $x$; the latter is the variance of the posterior mean of $V_T$.

\subsection{The auction and the location game}

A first-price auction allocates one right with payoff $V_T$. Each proposer either abstains or submits a finite bid $b_i\geq L$, where $L\in\R$ is the reserve. For the \emph{unrestricted no-reserve auction}, $L=-\infty$. The highest bid wins and is paid to the seller. Ties are broken uniformly among the highest bidders. If all proposers abstain, there is no allocation or payment. Proposers are risk-neutral: a winner bidding $b_i$ receives $V_T-b_i$, and a non-winner receives zero.

The unrestricted no-reserve auction is our baseline model. 
 A finite reserve, including a non-negativity restriction, changes the equilibrium; Section~\ref{sec:reserve} treats that change. A monopolist requires a finite reserve, since with no lower bid bound it has no finite optimal bid.

The game has two stages. Proposers first choose locations simultaneously, before observing the Brownian paths. They then observe \eqref{eq:information} and bid. An \emph{auction equilibrium} is a Bayesian Nash equilibrium of this second stage at fixed locations. For a specified selection of auction equilibria, let $\Pi_i(x_i,x_{-i})$ be proposer $i$'s ex ante expected auction profit. A \emph{location equilibrium} is a Nash equilibrium of the resulting complete-information game with payoffs $\Pi_i$.

With two proposers, the no-reserve continuation used throughout is the monotone equilibrium with zero limiting rent at the lowest signal quantile, defined in Section~\ref{sec:general-auction}. We focus on this equilibrium because it extends the symmetric equilibrium of Milgrom and Weber~\cite{milgromweber} to asymmetric signals, and coincides with it when the two signals are identically distributed. The unrestricted auction has other equilibria (Remark~\ref{rem:selection}), and some of them can support co-location; our location predictions are for the selected equilibrium. At profiles with no exclusive information for one proposer, we give an actual mixed continuation, rather than defining payoffs only by a limit. For more than two proposers, we will state the required continuation assumptions separately.

\subsection{Interpretation}
\label{sec:interpretation}
The bidders are proposers operated together with searchers or builders that pursue the same
financial opportunity (or operated independently but with pass-through). For example, an on-chain application can auction the right to execute a
fixed trade. The winner pays its bid to the application or the user supplying the opportunity and
receives the trade's payoff at settlement. Thus $V_T$ is the monetary value of winning before the
auction payment, not the proposer reward or necessarily the external price itself. An external
reference price can determine this value, for example through the mark-to-market payoff of the trade.
The Brownian specification is a stylized model of this value's innovations.

\begin{table}[t]
\caption{Model primitives and their interpretation.}
\label{tab:interpretation}
\centering
\small
\begin{tabular}{@{}p{0.29\linewidth}p{0.66\linewidth}@{}}
\toprule
Model & Interpretation \\
\midrule
$V_T$ & Monetary payoff of the auctioned opportunity before the bid payment \\
Source $k$, variance $w_k$ & A trading venue or other source of payoff-relevant information \\
$d_k(x)$ & Delay with which source information is available for bidding at $x$ \\
$T$ & Settlement time and baseline bidding deadline \\
Proposer & A proposer with an integrated searcher or builder \\
First-price auction & An on-chain sale of the opportunity to the highest included bidder \\
\bottomrule
\end{tabular}
\end{table}

The number of proposers is the protocol-design choice of interest. A single proposer controls
which bids enter its block and can exclude every competing bid~\cite{fox2023censorship}. It
therefore acquires the right at the reserve whenever the right is worth more, which is the
one-proposer model of Section~\ref{sec:monopoly}. Multiple concurrent proposers each include
their own bids, so the right is contested and the auction of Section~\ref{sec:two} applies.

The model abstracts from consensus, from network topology beyond pairwise delays, from collusion
among proposers, and from the other tasks involved in building a block. Validators enter only
through the bidding deadline. Section~\ref{sec:deadlines} lets that deadline depend on location,
for example through the placement of the attesters who must receive a proposer's block.

\section{One proposer}
\label{sec:monopoly}

A monopolist's posterior mean at $x$ is
\[
 \mu(x):=\E[V_T\mid\Fscr_1(x)]
 =V_0+\sum_k\sigma_k B^k_{T-d_k(x)},
\]
and its posterior error has variance $D(x)$. Bidding is trivial, its best response is to pay the reserve $L$ when $\mu(x)\geq L$ and to abstain otherwise. Consequently, its ex ante profit is
\begin{equation}
 \Pi_1(x)=\E[(\mu(x)-L)_+].
 \label{eq:monopoly-profit}
\end{equation}

More information increases the dispersion of the posterior mean without changing its expectation. Since the proposer can abstain after an unfavorable signal, the resulting increase in the value of the positive part is beneficial. In the Brownian model, this gives a particularly simple location criterion.

\begin{restatable}[Monopolist location]{theorem}{restatethmmonopoly}
\label{thm:monopoly}
The monopolist's optimal locations are exactly those that minimize variance-weighted delays
$
 \argmin_{x\in\Xscr}D(x).$
\end{restatable}

\begin{proof}
Write $s=\sqrt{I(x)}$. For $s>0$, normal integration gives
\[
 \Pi_1(x)=(V_0-L)\Phi\!\left(\frac{V_0-L}{s}\right)
             +s\phi\!\left(\frac{V_0-L}{s}\right),
\]
where $\phi$ and $\Phi$ are the standard normal density and distribution function. Its derivative with respect to $s$ is $\phi((V_0-L)/s)>0$. At $s=0$ the continuous extension is $(V_0-L)_+$, and every positive $s$ gives a strictly larger value. Thus profit is strictly increasing in $I(x)=TW-D(x)$. Compactness and continuity give existence of a minimizer of $D$.
\end{proof}

With Euclidean delays, the objective is $\sum_k w_k\norm{x-z_k}_2$. Its minimizers are the variance-weighted geometric medians, restricted to $\Xscr$ if necessary. On a feasible interval containing the source sites, a \emph{weighted median} is a point with at most half the total weight strictly to either side. In particular, if there is a source with $w_j\geq\sum_{i\neq j}w_i$, then co-locating with that source is optimal.

For two sources on a Euclidean segment, the proposer therefore locates at the higher-variance source when the weights differ. If they are equal, exactly the feasible points between the two sources are optimal. More generally, equality in the dominance condition guarantees optimality at the source, but not uniqueness.

\section{Two proposers}
\label{sec:two}

We first record an equilibrium of the auction with general signals and a general common-value aggregator, then specialize it to the Brownian model and use the resulting rents to solve the location game. The bidding construction survives general aggregation; the location potential needs the additive Brownian structure.

\subsection{A general auction equilibrium}
\label{sec:general-auction}

Fix common information $c$, observed by both proposers. Conditional on $c$, let $X_A$ and $X_B$ be independent scalar signals with continuous, strictly increasing quantile functions $t_A$ and $t_B$ on $0<q<1$. Write $F_i$ for the corresponding distribution function. Assume that the common terminal value is conditionally integrable. Its conditional expected value is
\[
 a(c,x,y):=\E[V\mid c,X_A=x,X_B=y].
\]
Assume that the function $a$ is continuous and nondecreasing in each signal.

All distributions and expectations in this general auction analysis are conditional on the fixed common information $c$. Suppress $c$ temporarily and define
\begin{equation}
 g(q,r):=a(c,t_A(q),t_B(r)),\qquad h(q):=g(q,q).
 \label{eq:quantile-values}
\end{equation}
Assume that $h$ is strictly increasing and $\int_0^1|h(q)|\,dq<\infty$. Monotonicity bounds $g(q,r)$ between $h(\min\{q,r\})$ and $h(\max\{q,r\})$, so all expected-value integrals below are finite.

A bidder's quantile is uniform, even when the two signal distributions differ. If both bidders use the same increasing bid as a function of their quantile, a report $r$ wins with probability $r$. At an equilibrium type $q$, the marginal value from increasing that report is $h(q)$. The first-order condition therefore suggests $(q\beta(q))'=h(q)$. Integrating from the bottom of the signal distribution yields the following equilibrium (the proof below verifies all putative deviations, not just the first-order condition).

Cheng and Tan~\cite{chengtan2010} give this quantile bidding construction for asymmetric first-price common-value auctions. We state and verify it here for the present signal assumptions and unrestricted bid space.

\begin{restatable}[Quantile equilibrium]{theorem}{restatethmquantile}
\label{thm:quantile}
In the unrestricted no-reserve auction, an equilibrium is
\begin{equation}
 b_i(c,x)=\beta(F_i(x);c),\qquad
 \beta(q;c):=\frac1q\int_0^q
 a(c,t_A(s),t_B(s))\,ds.
 \label{eq:general-bid}
\end{equation}
The quantile bid is continuous and strictly increasing. Each proposer wins with probability $1/2$. At quantile $q$, their expected profits are
\begin{align}
 u_A(q)&=\int_0^q[g(q,s)-g(s,s)]\,ds,\label{eq:general-rent-A}\\
 u_B(q)&=\int_0^q[g(s,q)-g(s,s)]\,ds.\label{eq:general-rent-B}
\end{align}
Both profits are nonnegative and tend to zero as $q\downarrow0$.
\end{restatable}

\begin{proof}
Continuity is immediate, and $q\beta'(q)=h(q)-\beta(q)>0$, since $h(q)$ exceeds its average over $0<s<q$. If proposer $A$ has quantile $q$ and bids as quantile $r$, its expected profit is
\[
 G_A(r;q)=\int_0^r g(q,s)\,ds-r\beta(r).
\]
Differentiating in $r$ and using $(r\beta(r))'=h(r)$ gives
\[
 \partial_rG_A(r;q)=g(q,r)-g(r,r).
\]
This is nonnegative for $r<q$ and nonpositive for $r>q$. Thus $r=q$ is a global best response. For proposer $B$, the derivative is $g(r,q)-g(r,r)$, with the same sign. If, in addition, $a$ is strictly increasing in the bidder's own signal, these inequalities are strict and $r=q$ is the unique best response.

A bid below the bid range loses. If the upper endpoint of the bid range is finite, bidding above it wins with certainty and costs at least as much as its limiting endpoint bid. Such a deviation cannot improve on the best quantile report. Abstention yields zero, while the displayed profit formulas are nonnegative by monotonicity. Finally,
\[
 0\leq u_i(q)\leq qh(q)-\int_0^q h(s)\,ds\longrightarrow0.
\]
The limit follows from monotonicity and integrability of $h$. Independent uniform quantiles give the winning probabilities.
\end{proof}

The proof does not use symmetry of $a$: it applies to any common, coordinatewise monotone aggregator with the stated diagonal properties. Independence, on the other hand, is substantive. With correlated signals, the probability of beating a reported quantile, conditional on the bidder's own quantile, need not equal that quantile, and \eqref{eq:general-bid} need not be an equilibrium. When the two signals are identically distributed, \eqref{eq:general-bid} is the symmetric equilibrium of Milgrom and Weber~\cite{milgromweber}.

\begin{remark}[The lower-tail boundary condition]
\label{rem:selection}
Among continuously differentiable increasing equilibria with positive signal densities, a common bid support, and zero limiting lowest-quantile rent, \eqref{eq:general-bid} is unique. The inverse first-order conditions match the two quantiles, after which $(q\beta(q))'=h(q)$ determines the bid up to an additive term $k/q$. The boundary condition selects $k=0$; Appendix~\ref{app:uniqueness} gives the argument and an example. Without it, $k<0$ also preserves the incentive comparisons and gives each lowest-quantile limit a positive rent $-k$, financed by arbitrarily negative bids. We do not claim uniqueness in this unrestricted class.
\end{remark}

\subsection{Additive values and information rents}
\label{sec:additive-auction}

For $a(c,x,y)=c+x+y$, the quantile formula separates across bidders. The resulting rent depends only on the bidder's own signal distribution.

\begin{restatable}[Additive signal rents]{proposition}{restatepropadditiverents}
\label{prop:additive-rents}
Suppose $X_A,X_B$ are independent and integrable, with the quantile regularity above, and $\E[V\mid c,X_A,X_B]=c+X_A+X_B$. Then
\begin{equation}
 \beta(q;c)=c+\frac1q\int_0^q[t_A(s)+t_B(s)]\,ds,
 \qquad
 \pi_i(c)=\frac12\E|X_i-X_i'|,
 \label{eq:additive-rents}
\end{equation}
where $\pi_i(c)$ is profit averaged over the private signals and $X_i'$ is an independent copy of $X_i$.
\end{restatable}

\begin{proof}
Substitution in \eqref{eq:general-rent-A}--\eqref{eq:general-rent-B} gives
$u_i(q)=qt_i(q)-\int_0^q t_i(s)\,ds$. Integrating over $q$ and changing the order of integration gives
\[
 \pi_i(c)=\int_0^1(2q-1)t_i(q)\,dq
 =\int_0^1\int_0^q[t_i(q)-t_i(s)]\,ds\,dq.
\]
The last integral is half the expected absolute difference of two independent draws.
\end{proof}

Neither the quantile equilibrium nor this rent formula requires normality. For example,
if $X_i$ is uniform on $-a_i<x<a_i$, then
$\beta(q;c)=c+(a_A+a_B)(q-1)$ and $\pi_i(c)=a_i/3$. For the geographic analysis, however, the signal distribution comes from observation delays, so we now return to \eqref{eq:value}.

\subsection{The Brownian signal decomposition}
\label{sec:decomposition}

Fix locations $x_A,x_B$, and abbreviate $d_{ki}:=d_k(x_i)$. The part of source $k$ observed by both proposers ends at $T-\max\{d_{kA},d_{kB}\}$. The common component is therefore
\begin{equation}
 C:=V_0+\sum_k\sigma_k B^k_{T-\max\{d_{kA},d_{kB}\}}.
 \label{eq:common}
\end{equation}
For sources where $A$ is faster, let $X_A$ be the sum of the increments from this common observation time to $T-d_{kA}$. Define $X_B$ in the same way on sources where $B$ is faster. Finally, let $\varepsilon$ be the sum of the increments after the more recent of the two observations. Thus
\begin{equation}
 V_T=C+X_A+X_B+\varepsilon.
 \label{eq:decomposition}
\end{equation}
These four components are independent. Their variances are
\begin{align}
 \Omega_A&:=\sum_k w_k[d_{kB}-d_{kA}]_+,&
 \Omega_B&:=\sum_k w_k[d_{kA}-d_{kB}]_+,\label{eq:exclusive}\\
 \Omega_0&:=\sum_k w_k\min\{d_{kA},d_{kB}\},&
 \Gamma&:=\sum_k w_k(T-\max\{d_{kA},d_{kB}\}).\label{eq:residual}
\end{align}
In particular, $C\sim N(V_0,\Gamma)$, $X_i\sim N(0,\Omega_i)$, and $\varepsilon\sim N(0,\Omega_0)$, with
\begin{equation}
 \Gamma+\Omega_A+\Omega_B+\Omega_0=TW.
 \label{eq:variance-accounting}
\end{equation}
Independence follows source by source from disjoint Brownian increments, and then across sources. All of $C$ is observed by both proposers because each observes the relevant source histories. Conditional on their full information, the remaining variables relevant for bidding are exactly $C$ and their own $X_i$; the additional observed path details do not change the conditional law of the opponent's increment or of $\varepsilon$.

\begin{figure}[tbp]
\centering
\begin{tikzpicture}[x=1.4cm,y=1.4cm,font=\small]
  \definecolor{spatialA}{RGB}{0,114,178}
  \definecolor{spatialB}{RGB}{213,94,0}
  \pgfmathsetmacro{\nearradius}{sqrt(0.4^2+0.45^2)}
  \pgfmathsetmacro{\farradius}{sqrt(1.6^2+0.45^2)}
  \coordinate (sourcezero) at (-1,0);
  \coordinate (sourceone) at (1,0);
  \coordinate (proposerA) at (-0.6,0.45);
  \coordinate (proposerB) at (0.6,-0.45);

  % Separate annuli keep each source's categories disjoint before overlay.
  \foreach \source/\shade in {sourcezero/spatialA,sourceone/spatialB} {
    \fill[black,fill opacity=0.07,even odd rule]
      (\source) circle[radius=2] (\source) circle[radius=\farradius];
    \fill[\shade,fill opacity=0.12,even odd rule]
      (\source) circle[radius=\farradius] (\source) circle[radius=\nearradius];
    \fill[pattern=north east lines,pattern color=black!25]
      (\source) circle[radius=\nearradius];
  }
  % Solid circles belong to source 0; dashed circles belong to source 1.
  \foreach \radius in {\nearradius,\farradius,2} {
    \draw[black!65,thin] (sourcezero) circle[radius=\radius];
    \draw[black!65,thin,dashed] (sourceone) circle[radius=\radius];
  }
  \foreach \source in {sourcezero,sourceone} {
    \draw[black!35,thin] (\source) -- (proposerA);
    \draw[black!35,thin] (\source) -- (proposerB);
  }

  \node[fill=white,inner sep=1pt] at (-2.10,1.47) {$C$};
  \node[fill=white,inner sep=1pt] at (2.10,1.47) {$C$};
  \node[text=spatialA!80!black] at (-2.10,-0.50) {$X_A$};
  \node[text=spatialB!80!black] at (2.10,0.50) {$X_B$};
  \node[fill=white,inner sep=1pt] at (-1,0.32) {$\varepsilon$};
  \node[fill=white,inner sep=1pt] at (1,-0.32) {$\varepsilon$};
  \filldraw[fill=black,draw=white,line width=0.5pt]
    (sourcezero) circle[radius=0.045];
  \filldraw[fill=black,draw=white,line width=0.5pt]
    (sourceone) circle[radius=0.045];
  \node[anchor=north,fill=white,inner sep=1pt] at (-1,-0.09) {$z_0$};
  \node[anchor=south,fill=white,inner sep=1pt] at (1,0.09) {$z_1$};
  \node[rectangle,draw=spatialA,fill=white,minimum size=4.5mm,inner sep=1pt]
    at (proposerA) {$A$};
  \node[rectangle,draw=spatialB,fill=white,minimum size=4.5mm,inner sep=1pt]
    at (proposerB) {$B$};

  %\node at (0,2.28) {Propagation radius $r=T-t$, with $T=2$};
  \filldraw[fill=black!7,draw=black!50] (-2.85,-2.40) rectangle (-2.55,-2.20);
  \node[anchor=west] at (-2.45,-2.30) {Common: $C$};
  \filldraw[fill=spatialA!12,draw=spatialA] (0.15,-2.40) rectangle (0.45,-2.20);
  \node[anchor=west] at (0.55,-2.30) {$A$ only: $X_A$};
  \filldraw[fill=white,draw=black!50,pattern=north east lines,pattern color=black!25]
    (-2.85,-2.80) rectangle (-2.55,-2.60);
  \node[anchor=west] at (-2.45,-2.70) {Neither: $\varepsilon$};
  \filldraw[fill=spatialB!12,draw=spatialB] (0.15,-2.80) rectangle (0.45,-2.60);
  \node[anchor=west] at (0.55,-2.70) {$B$ only: $X_B$};
\end{tikzpicture}
\caption{Spatial information decomposition.
Sources are at $z_0,z_1$; proposers are at
$x_A,x_B$.
Each radius $r=T-t$ represents an update emitted at source time $t$.
The inner disk contains updates that have reached neither proposer;
the middle annulus contains updates that have reached only the nearer proposer;
the outer annulus contains older updates that have reached both.
Solid circles belong to $z_0$ and dashed circles to $z_1$.
Labels identify contributions to the value decomposition.
Overlap between different sources' bands is purely geometric;
radial widths represent time intervals.}
\label{fig:information-timeline}
\end{figure}
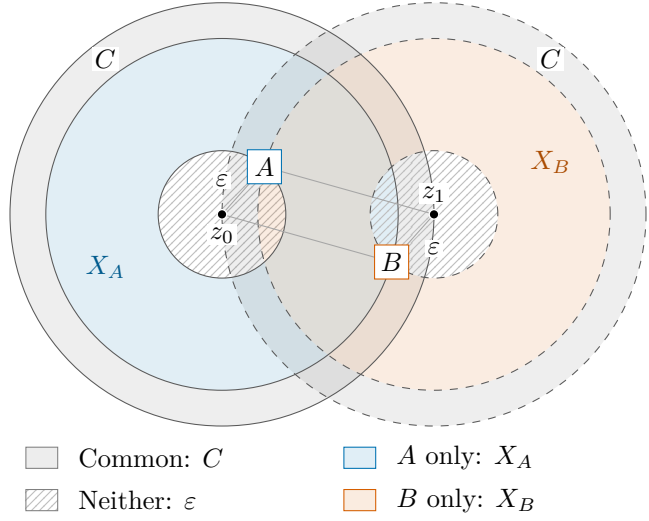

Figure~\ref{fig:information-timeline} illustrates how propagation distances separate common, exclusive, and unobserved information from two sources.
% Radius is the distance traveled by an update by settlement, r = T - t.
% Annular widths encode time intervals; their areas do not encode variance.

%%% Local Variables:
%%% mode: LaTeX
%%% TeX-master: t
%%% End:

Write $s_i:=\sqrt{\Omega_i}$ and $\lambda(z):=\phi(z)/\Phi(z)$. The normal quantile function gives an explicit bid and rent.

\begin{restatable}[Gaussian bids and profits]{corollary}{restatecorgaussian}
\label{cor:gaussian}
If $s_A,s_B>0$, the selected no-reserve equilibrium is
\begin{equation}
 b_i(C,X_i)=C-(s_A+s_B)\lambda(X_i/s_i).
 \label{eq:normal-bid}
\end{equation}
The higher standardized signal wins. At standardized signal $z$, proposer $i$'s interim profit is $s_i[z\Phi(z)+\phi(z)]$. Its ex ante profit and the seller's expected revenue are
\begin{equation}
 \Pi_i(x_A,x_B)=\frac{s_i}{\sqrt\pi},\qquad
 R(x_A,x_B)=V_0-\frac{s_A+s_B}{\sqrt\pi}.
 \label{eq:normal-rents}
\end{equation}
\end{restatable}

\begin{proof}
For $q=\Phi(z)$, integration of the normal quantile gives
$\int_0^q\Phi^{-1}(t)\,dt=-\phi(z)$. This proves the bid and interim profit formulas. Moreover, $X_i-X_i'\sim N(0,2s_i^2)$, so Proposition~\ref{prop:additive-rents} gives $\pi_i(c)=s_i/\sqrt\pi$ for every $c$. Averaging over $C$ gives the ex ante profit. There is always an allocation, and its value is common, so revenue equals expected value minus total expected bidder profit.
\end{proof}

A larger exclusive variance raises a proposer's rent, not its winning probability. At nondegenerate profiles the latter remains $1/2$. This distinction is important for location choice.

\begin{restatable}[Degenerate information]{proposition}{restatepropdegenerate}
\label{prop:degenerate}
Formula~\eqref{eq:normal-rents} extends to every location profile. If $s_A>0$ and $s_B=0$, proposer $A$ uses \eqref{eq:normal-bid}, while proposer $B$ draws an independent uniform quantile $Q$ and bids
\[
 C-s_A\lambda(\Phi^{-1}(Q)).
\]
This is an auction equilibrium, with profits $s_A/\sqrt\pi$ and zero. If $s_A=s_B=0$, both bid $C$ and receive zero profit.
\end{restatable}

\begin{proof}
In the first case the uninformed proposer reproduces the same bid distribution as a standard normal signal would generate. The informed proposer's deviation calculation is unchanged. At any bid corresponding to quantile $q$, the uninformed proposer's expected profit is
\[
 \int_0^q\bigl[C+s_A\Phi^{-1}(t)\bigr]dt
 -q\bigl[C-s_A\lambda(\Phi^{-1}(q))\bigr]=0.
\]
Bidding at or above $C$ has nonpositive expected profit; all remaining bids are covered by this identity or lose. If both exclusive variances are zero, conditional value is $C$: outbidding $C$ loses in expectation and underbidding cannot earn a positive profit.
\end{proof}

\subsection{The location game}
\label{sec:location}

The profit formula makes the location incentive transparent. Proposer $A$ gains only from being faster than $B$, and values a delay advantage at source $k$ in proportion to $w_k$. It does not benefit from observing information which its competitor already sees equally early. The two incentives are compatible with a common objective in the potential-game sense stated below.

\begin{restatable}[Location potential]{theorem}{restatethmpotential}
\label{thm:potential}
In the two-proposer no-reserve location game, $-\Omega_0$ is an ordinal potential. The pure location equilibria are exactly the profiles that minimize $\Omega_0$ globally in each coordinate separately. A pure location equilibrium exists, and every global minimizer of
\begin{equation}
 \min_{x_A,x_B\in\Xscr}\sum_k w_k\min\{d_k(x_A),d_k(x_B)\}
 \label{eq:two-median}
\end{equation}
is a location equilibrium.
\end{restatable}

\begin{proof}
The identities $[u-v]_+=u-\min\{u,v\}$ give
\begin{equation}
 \Omega_A=D(x_B)-\Omega_0,\qquad
 \Omega_B=D(x_A)-\Omega_0.
 \label{eq:potential-identities}
\end{equation}
With the opponent's location fixed, a proposer's exclusive variance increases exactly when $\Omega_0$ decreases, and by \eqref{eq:normal-rents} the same is true of its profit. Thus $-\Omega_0$ is an ordinal potential, while it is an exact potential for the variance payoffs $\Omega_i$. The best-response characterization follows. The continuous function $\Omega_0$ has a global minimizer on the compact product space; no unilateral change can improve a coordinate there.
\end{proof}

The distinction between an exact and an ordinal potential follows the terminology of Monderer and Shapley~\cite{monderershapley}. Here the square root preserves the sign of a payoff change, not its magnitude. Also, a coordinatewise minimizer need not be a global minimizer. The theorem does not identify all location equilibria with solutions of the joint two-median problem \eqref{eq:two-median}.

\begin{restatable}[Co-location in equilibrium is non-generic]{corollary}{restatecorcolocation}
\label{cor:colocation}
A co-located profile $(z,z)$ is a location equilibrium if and only if $z$ minimizes the delay to every information source,
$
 z\in\bigcap_{k=1}^K\argmin_{x\in\Xscr}d_k(x).$
Hence a co-location equilibrium exists if and only if this intersection is nonempty.
\end{restatable}

\begin{proof}
At co-location both profits are zero. A deviation to $x$ yields exclusive variance $\sum_k w_k[d_k(z)-d_k(x)]_+$. This is zero for every $x$ exactly when $z$ minimizes each positive-weight source delay.
\end{proof}

For distance delays and at least two distinct source sites available, no location minimizes delay
to all information sources simultaneously. Equal locations therefore cannot be a two-proposer
equilibrium. This conclusion does not depend on either source being sufficiently important. The
following provides an example:
\begin{restatable}[Two sources on a line]{proposition}{restateproptwosources}
  \label{prop:two-sources}
  Let $\Xscr=[0,1]$, with sources at $0$ and $1$, weights $w_0,w_1>0$, and distance delays. Up to exchanging the proposers, the unique pure location equilibrium is $(x_A,x_B)=(0,1)$.
\end{restatable}

\begin{proof}
Co-location is excluded by Corollary~\ref{cor:colocation}. Relabel the proposers so that $a:=x_A<x_B=:b$. Then
\[
 \Omega_A=w_0(b-a),\qquad \Omega_B=w_1(b-a).
\]
If $a>0$, moving $A$ to $0$ strictly increases its profit. If $b<1$, moving $B$ to $1$ does the same. Conversely, against a proposer at $1$, exclusive variance is $w_0(1-x)$ and is maximized at $0$; the other best response is symmetric.
\end{proof}

\label{app:line-split}

With more sources on a line, a global solution of the potential problem is still easy to describe. Order the distinct source sites as $z_1<\cdots<z_K$, and suppose the feasible interval contains them. For a consecutive nonempty group define
\[
 J(r,t):=\min_y\sum_{k=r}^t w_k|y-z_k|.
\]
Any weighted median of that group attains the minimum.

\begin{proposition}[A location equilibrium from a split]
\label{prop:line-split}
For $K\geq2$, minimize $J(1,j)+J(j+1,K)$ over $1\leq j<K$. Weighted medians of the two groups at a minimizing split solve \eqref{eq:two-median}, and hence form a location equilibrium.
\end{proposition}

\begin{proof}
Two ordered facilities assign their nearest sources to consecutive groups, breaking a midpoint tie in either direction. Their total cost is at least the sum of the two groups' optimal one-facility costs. Conversely, placing facilities at weighted medians of a split has nearest-facility cost at most the split's assigned cost. These inequalities identify the two optima. Empty groups cannot be needed for a global optimum with two distinct positive-weight sources: a single facility leaves some source at positive distance, and placing the unused facility there strictly lowers the cost.
\end{proof}

Separation is costly for the seller. By \eqref{eq:normal-rents}, revenue is $V_0-(s_A+s_B)/\sqrt\pi$, so the exclusive information that separates the proposers is paid for out of the seller's revenue, which is highest, at $V_0$, when the proposers see the same information. Whoever collects the auction revenue therefore prefers the concentrated geography that competition works against. Section~\ref{sec:deadlines} shows one way to get it: a single fixed auctioneer makes co-location an equilibrium.

\subsection{A finite reserve}
\label{sec:reserve}

A reserve changes the lower boundary of the auction. In particular, unequal signal distributions can make one proposer enter earlier and pool at the reserve while the other abstains. Simply truncating the no-reserve bid at $L$ does not in general give an equilibrium. We first describe this boundary for the general aggregator, and then specialize to Gaussian signals.

Return to \eqref{eq:quantile-values}, conditional on common information $c$, and now suppose the aggregator is strictly increasing in each signal. The expected values from entering against all opponent quantiles below $q$ are
\[
 e_A(q):=\frac1q\int_0^q g(q,s)\,ds,\qquad
 e_B(q):=\frac1q\int_0^q g(s,q)\,ds.
\]
Both are continuous and strictly increasing. Suppose there is $0<q_R<1$ such that
\begin{equation}
 \min\{e_A(q_R),e_B(q_R)\}=L.
 \label{eq:general-entry}
\end{equation}
It is unique. Call a proposer attaining the minimum $W$ and the other $H$, and reorder the arguments of $g$ so that the $H$ quantile comes first. Thus
$q_R^{-1}\int_0^{q_R}g(s,q_R)\,ds=L$, while the analogous entry value for $H$ is at least $L$. Define
\begin{equation}
 p_R:=\inf\left\{p\in(0,q_R]:
       \frac1{q_R}\int_0^{q_R}g(p,s)\,ds\geq L\right\}.
 \label{eq:general-pool-cutoff}
\end{equation}
The set is nonempty. If $p_R>0$, equality holds at $p_R$; if $p_R=0$, all lower quantiles of $H$ are willing to enter. Labels $H,W$ describe entry incentives here, not a general ordering by variance.

\begin{theorem}[Reserve equilibrium]
\label{thm:reserve-general}
Under \eqref{eq:general-entry}, an auction equilibrium has the following form. Proposer $W$ abstains below $q_R$. Proposer $H$ abstains below $p_R$ and bids $L$ on $p_R\leq q\leq q_R$. Above $q_R$, both use
\begin{equation}
 \beta_L(q;c)=\frac{q_RL+\int_{q_R}^q h(s)\,ds}{q},
 \label{eq:reserve-general-bid}
\end{equation}
which is strictly increasing from $L$. The probabilities of no sale and of a sale at the reserve are respectively $p_Rq_R$ and $(q_R-p_R)q_R$.
\end{theorem}

The proof is in Appendix~\ref{app:reserve-proof}. Above the reserve, the winning quantile has unconditional density $2q$ for $q>q_R$. Conditional seller revenue is consequently
\[
 R_L(c)=L(q_R-p_R)q_R+2\int_{q_R}^1q\beta_L(q;c)\,dq.
\]
The higher part of the bid solves the same differential equation as before. The new work is to verify entry and deviations to the reserve itself, where proposer $H$ can have an atom. Condition~\eqref{eq:general-entry} specifies a two-active-bidder regime. With bounded signals and sufficiently high reserves it need not hold. In the additive Gaussian model below, with both exclusive variances positive, it holds for every finite reserve.

For the additive Gaussian model, relabel so that $s_H\geq s_W>0$. Here $e_H(q)-e_W(q)=(s_H-s_W)(z+\lambda(z))\geq0$, where $z=\Phi^{-1}(q)$, so the entry ordering agrees with the variance ordering. Let $z_R$ be the unique solution of
\begin{equation}
 L-c=s_Wz_R-s_H\lambda(z_R),\qquad q_R:=\Phi(z_R),
 \label{eq:normal-entry}
\end{equation}
and define
\begin{equation}
 z_H^0:=\frac{L-c+s_W\lambda(z_R)}{s_H},\qquad
 p_R:=\Phi(z_H^0).
 \label{eq:normal-pool-cutoff}
\end{equation}
Then $z_H^0\leq z_R$. A weak proposer with standardized signal at most $z_R$ abstains. A strong proposer abstains below $z_H^0$, bids $L$ between $z_H^0$ and $z_R$, and otherwise uses the common increasing branch
\begin{equation}
 \beta_L(q;c)=c+
 \frac{q_R(L-c)+(s_H+s_W)
       [\phi(z_R)-\phi(\Phi^{-1}(q))]}{q}.
 \label{eq:normal-reserve-bid}
\end{equation}
The atom at the reserve is strictly positive exactly when $s_H>s_W$. It resolves the different entry indifference conditions; it is not an arbitrary choice of equilibrium behavior.

For the profit formulas define two normal integrals,
\[
 A(z):=\int_z^\infty\Phi(t)\Phi(-t)\,dt,
 \qquad B(u,z):=\int_u^z\Phi(-t)\,dt.
\]

\begin{proposition}[Gaussian reserve profits]
\label{prop:reserve-profits}
Conditional on $C=c$, expected profits in the reserve equilibrium are
\begin{equation}
 \pi_W(c)=s_W A(z_R),\qquad
 \pi_H(c)=s_H\bigl[A(z_R)+q_R B(z_H^0,z_R)\bigr].
 \label{eq:reserve-profits}
\end{equation}
As $L-c\to-\infty$, the bids at every fixed interior quantile converge to \eqref{eq:normal-bid}, and profits converge to $s_i/\sqrt\pi$.
\end{proposition}

\begin{proof}
The function $s_Wz-s_H\lambda(z)$ is strictly increasing from $-\infty$ to $+\infty$, since $\lambda'(z)=-\lambda(z)(z+\lambda(z))<0$. This proves existence and uniqueness in \eqref{eq:normal-entry}. Moreover,
\[
 z_H^0-z_R=(s_W/s_H-1)(z_R+\lambda(z_R))\leq0.
\]
The two entry conditions and normal quantile integration give \eqref{eq:normal-reserve-bid} from Theorem~\ref{thm:reserve-general}.

The allocation probability of an active $W$ type $z$ is $\Phi(z)$, while that of $H$ is $q_R$ in the pooling interval and $\Phi(z)$ above it. The derivative of interim profit with respect to the standardized signal is $s_i$ times this allocation probability. Profit is zero at the respective entry cutoff. Integrating this envelope, and then integrating over the bidder's normal type, gives \eqref{eq:reserve-profits}. As $L-c\to-\infty$, $z_R\to-\infty$, $A(z_R)\to1/\sqrt\pi$, and the pooling term tends to zero. Formula~\eqref{eq:normal-reserve-bid} gives the bid limit.
\end{proof}

For a fixed nominal reserve, location payoffs are obtained by averaging \eqref{eq:reserve-profits} over $C\sim N(V_0,\Gamma)$:
\begin{equation}
 \Pi_i^L(x_A,x_B)=\E_C[\pi_i(C)].
 \label{eq:reserve-location}
\end{equation}
Thus both exclusive variances and the distribution of common information matter. If instead the reserve is indexed to common information, $L(C)=C+\ell$, the conditional cutoffs depend on the fixed increment $\ell$, and this averaging is unnecessary. Such indexing presumes that the seller can observe or implement $C$.

Unlike \eqref{eq:normal-rents}, reserve profits need not depend only on a proposer's own exclusive variance. The potential proof therefore does not extend merely by replacing the bid with \eqref{eq:normal-reserve-bid}. Equation~\eqref{eq:reserve-location} specifies the reserve location problem, but no general two-median characterization is available. Appendix~\ref{app:reserve-degenerate} gives the auction continuations when an exclusive variance is zero, so that the reserve payoffs are also defined at those profiles.

\draftpagebreak

\section{More than two proposers}
\label{sec:many}

With two proposers, every source increment is common, exclusive to one proposer, or unobserved by both. With more proposers, an increment may be observed by any intermediate subset. The resulting information is generally neither independent across proposers nor summarized by one private scalar per proposer. We therefore cannot extend Theorem~\ref{thm:quantile} by replacing two uniform quantiles with $M$ of them.

There is nevertheless a direct argument against duplication of information. Its two parts are distinct. A proposer with an increment unknown to every rival can obtain a positive profit. Proposers with identical information receive zero profit under regular pure bidding continuations. A location change that replaces duplicated information by an exclusive increment then gives a profitable deviation. We state the conditions on both parts explicitly.

For a profile of locations define the variance of information exclusive to proposer $i$ relative to \emph{all} its rivals by
\begin{equation}
 \Omega_i^{\rm ex}(x_i,x_{-i})
 :=\sum_k w_k\left[\min_{j\ne i}d_k(x_j)-d_k(x_i)\right]_+.
 \label{eq:many-exclusive}
\end{equation}
If $\Omega_i^{\rm ex}>0$, proposer $i$ can earn a strictly positive profit against any profile of finite rival bids, with or without a reserve, by bidding only after a high exclusive increment (\Cref{lem:exclusive-profit}). 
This needs neither an explicit $M$-bidder equilibrium nor a guess about its allocation rule: any equilibrium at such a profile must give the proposer a positive profit.

Consider the group of all proposers at a location occupied by at least two of them. They observe the same history. Call their continuation \emph{regular and pure at this profile} if their bids are deterministic functions of that history and, whenever some member bids, the highest outsider bid conditional on that history has no atom at their highest submitted bid. Conditional expected payoffs are assumed finite. If the group contains every proposer, the outsider condition is vacuous. The condition rules out a discontinuous change in the set of outsider types defeated by an arbitrarily small overbid. Under it, every member of a co-located group earns zero expected profit, since any member could otherwise overbid its tied partners (\Cref{lem:duplicate-zero}).

\begin{lemma}[A strictly private increment]
\label{lem:exclusive-profit}
If $\Omega_i^{\rm ex}>0$, proposer $i$ can earn a strictly positive ex ante profit against any profile of finite rival bids with integrable terminal value. The assertion holds with no reserve and with any finite reserve.
\end{lemma}
\begin{proof}
Write $V_T=X+Y$, where $X$ is the sum of the increments exclusive to $i$. Then $X\sim N(0,\Omega_i^{\rm ex})$ is observed by $i$ and is independent of both $Y$ and all rival information and private randomizations. Let $B_{-i}$ be the largest rival bid, setting it to $-\infty$ if all rivals abstain. Choose a finite $b$, at least the reserve when there is one, such that $p:=\PR(B_{-i}<b)>0$. Such a $b$ exists because the rival bids are finite. Put $m:=\E[Y\mid B_{-i}<b]$.

Proposer $i$ can bid $b$ only when $X\geq t$ and abstain otherwise. Choose $t>b-m$. Its expected profit on strict wins is
\[
 p\,\E[(X+m-b)\I{X\geq t}]>0.
\]
To avoid any additional ties, choose $b$ outside the at most countable set of atoms of $B_{-i}$; this can be done while retaining $p>0$. The normal upper tail has positive probability for every finite $t$, proving the claim.
\end{proof}

\begin{lemma}[Zero rent from duplicated information]
\label{lem:duplicate-zero}
In an auction equilibrium with a regular pure continuation for a co-located group, every member of that group has zero expected profit, conditional on its common information.
\end{lemma}
\begin{proof}
Fix the common history and suppose some group member submits a bid. Let $b$ be the highest group bid, and let $m$ members submit it. Let $A$ be the conditional expected profit from winning against outsiders at price $b$. Only these $m$ members can win, and each receives $A/m$. Abstention implies $A\geq0$.

If $A>0$ and $m\geq2$, one of the tied members can bid $b+\epsilon$ and obtain a profit tending to $A$ as $\epsilon\downarrow0$, instead of $A/m$. If $m=1$, another member of the group can make the same deviation from its zero profit. The no-atom condition and conditional integrability justify this limit. Both cases contradict equilibrium. Thus $A=0$. If the entire group abstains, all its profits are already zero.
\end{proof}

\begin{restatable}[No co-location]{theorem}{restatethmmanydistinct}
\label{thm:many-distinct}
Suppose $K\geq M$, delays are distances to distinct positive-weight source sites, and all source sites are feasible locations. Then every pure location equilibrium whose co-located groups have regular pure auction continuations has pairwise distinct proposer locations.
\end{restatable}

\begin{proof}
Suppose instead that some proposers are co-located. Then at most $M-1$ distinct locations are occupied, so $K\geq M$ implies that some source site is unoccupied. Members of the co-located group receive zero profit by Lemma~\ref{lem:duplicate-zero}. If one of them moves to the unoccupied source site, its delay to that source becomes zero while every rival has strictly positive delay. Equation~\eqref{eq:many-exclusive} is therefore positive, and Lemma~\ref{lem:exclusive-profit} gives a strictly profitable strategy against any finite-bid continuation of the rivals. This contradicts location optimality.
\end{proof}

In particular, when delays are distances to at least two distinct positive-weight source sites that are feasible locations, \emph{full} co-location is excluded for any $M\geq2$ whose continuation there is pure: at least one source is vacant, and the outsider condition is automatic. The pairwise conclusion for arbitrary profiles requires the additional conditions in Theorem~\ref{thm:many-distinct}. Neither the existence of a general $M$-bidder auction equilibrium nor the existence of a pure location equilibrium for that general game has been established here. Distinct locations also need not be far apart: the theorem rules out duplication, not proximity, and gives no lower bound on the distance between proposers.

The qualifications matter. With one source and $M\geq3$, all proposers can co-locate at the source in equilibrium, since a proposer who leaves still faces two fully informed rivals (Appendix~\ref{app:one-source-many}). And \Cref{lem:duplicate-zero} does not cover mixed bidding with unbounded subsidies.
\section{Location-dependent delays and deadlines}
\label{sec:deadlines}

Information collection is not the only source of latency. A proposer may also need time to deliver its bid, payload, or other message before a common deadline. We model this by an additional delay $\tau(x)$ depending on its location. If source propagation takes $\delta_k(x)$ and the decision must be made by $T-\tau(x)$, the effective source delay is
\begin{equation}
 d_k(x)=\delta_k(x)+\tau(x).
 \label{eq:effective-delay}
\end{equation}
We assume throughout this section that these effective delays are at most $T$.

\subsection{An attester-dependent deadline}

A simple specification uses attesters at $a_1,\ldots,a_n$, with positive weights $v_1,\ldots,v_n$. To reach a fraction $0<\gamma\leq1$ of total attester weight, define the delivery time
\begin{equation}
 \tau_\gamma(x):=\inf\left\{t\geq0:
 \sum_{j=1}^n v_j\I{\rho(x,a_j)\leq t}
 \geq\gamma\sum_{j=1}^n v_j\right\}.
 \label{eq:quorum}
\end{equation}
For metric communication times, $\tau_\gamma$ is continuous: if $x$ changes to $y$, every delivery time changes by at most $\rho(x,y)$, and the same bound holds for the weighted order statistic.

\begin{restatable}[Effective-delay reduction]{proposition}{restatepropeffectivedelay}
\label{prop:effective-delay}
Under \eqref{eq:effective-delay}, the monopolist minimizes
\[
 \sum_k w_k\delta_k(x)+W\tau(x).
\]
The two-proposer no-reserve game has ordinal potential
\[
 -\sum_k w_k\min\{\delta_k(x_A)+\tau(x_A),
                       \delta_k(x_B)+\tau(x_B)\}.
\]
\end{restatable}

\begin{proof}
The last observed source time is $T-d_k(x)$, so the information structure is exactly \eqref{eq:information} with the effective delay. Apply Theorems~\ref{thm:monopoly} and \ref{thm:potential}.
\end{proof}

A constant delivery delay cancels from the exclusive variances and therefore does not change two-proposer no-reserve location incentives. It still changes common and residual information, and can change reserve payoffs and price informativeness. Location-dependent delivery delays do not cancel.

For a fixed assignment of sources $S$ to a proposer, the assigned part of the potential cost is
\begin{equation}
 \sum_{k\in S}w_k\delta_k(x)+W_S\tau(x),
 \qquad W_S:=\sum_{k\in S}w_k.
 \label{eq:cluster-deadline}
\end{equation}
Thus delivery latency is weighted by the variance of the sources served. This is a conditional, fixed-assignment objective. If movement changes which proposer is faster, the assignment must also be recomputed; an unconstrained minimizer of \eqref{eq:cluster-deadline} need not be a best response in the full location game.

\subsection{A fixed auctioneer}

Suppose source information and bids propagate according to the same communication-time metric $\rho$, and bids must reach an auctioneer at $z\in\Xscr$ by time $T$. Thus
\begin{equation}
 \delta_k(x)=\rho(z_k,x),\qquad
 \tau(x)=\rho(x,z).
 \label{eq:auctioneer-delay}
\end{equation}
A proposer at $x$ must send its bid by $T-\rho(x,z)$. For a fixed assignment of sources, a proposer's cost is then a weighted geometric-median problem with the auctioneer as an additional point, weighted by the variance of the sources it serves (\eqref{eq:cluster-deadline}). Moving from the auctioneer toward a source can reduce the time needed to receive its updates, but adds a bid-delivery delay.

\begin{restatable}[Auctioneer co-location and equilibrium selection]{proposition}{restatepropauctioneer}
\label{prop:auctioneer}
Bidders minimize their weighted information delay by locating at the auctioneer location $z$. Every profile $(z,y)$ or $(y,z)$, with $y\in\Xscr$, is a two-proposer no-reserve location equilibrium. In particular, $(z,z)$ is an equilibrium.
\end{restatable}

\begin{proof}
The triangle inequality gives
\[
 d_k(x)=\rho(z_k,x)+\rho(x,z)
 \geq\rho(z_k,z)=d_k(z).
\]
Against a proposer at $z$, any location has zero exclusive variance and is a best response. Against $y$, locating at $z$ maximizes every term $[d_k(y)-d_k(x)]_+$, so is also a best response. Interchanging the proposers gives the same conclusion for $(y,z)$.
\end{proof}

The auctioneer makes co-location sustainable but does not force it: against a rival at $z$, a proposer cannot gain an information advantage anywhere. At $(z,z)$ all information rents vanish; at $(z,y)$ the proposer at $z$ has exclusive variance $D(y)-D(z)$.

The minimizer of $d_k$ need not be unique. Equality $d_k(x)=d_k(z)$ holds precisely when $\rho(z_k,x)+\rho(x,z)=\rho(z_k,z)$. In particular, when shortest paths exist, every feasible point on a shortest path from source $z_k$ to the auctioneer has the same effective delay for that source.

The auctioneer's own location also matters for revenue. With proposer locations fixed, revenue depends on the auctioneer's location only through the difference between the two proposers' delivery delays, and a revenue-maximizing location can be found among finitely many candidates (\Cref{prop:auctioneer-revenue} in Appendix~\ref{app:more-location}). An auctioneer that commits to a location before the proposers choose theirs cannot, however, guarantee itself the co-location revenue $V_0$ without an equilibrium-selection argument.

\section{Price informativeness}
\label{sec:informativeness}
In this section, we study how informative auction outcomes are for an outside observer who wants to infer the terminal value.  An observer who sees only the auction outcome must also infer how information affected the bids. We measure the resulting error about the terminal value, rather than treating the winning bid itself as an unbiased estimate.

\subsection{The observer's information and the monopoly benchmark}

Fix a public observation time $0\leq t_{\rm pub}<T$ and let $H:=T-t_{\rm pub}$. The outside observer receives a signal delayed by $H$,\footnote{An observer of the undelayed signal would know the terminal value $V_T$ and learn nothing from the auction outcome.} so that, besides the auction outcome, the observer knows
\[
 c_{\rm pub}:=V_{t_{\rm pub}}
 =V_0+\sum_k\sigma_kB^k_{t_{\rm pub}}.
\] Moreover, the observer learns the \emph{winning bid} and the \emph{winner's identity} in the auction. Assume $d_k(x)\leq H$ for every feasible location and every source, i.e. the outside observer is worse informed than the auction participants.

Condition on $c_{\rm pub}$. The remaining prior variance of $V_T$ is $HW$. The decomposition in Section~\ref{sec:decomposition} continues to hold, except that $C$ now has conditional mean $c_{\rm pub}$ and conditional variance
\begin{equation}
 \Gamma=\sum_k w_k(H-\max\{d_{kA},d_{kB}\}).
 \label{eq:public-common-var}
\end{equation}
The exclusive and residual variances are unchanged, and their sum with $\Gamma$ is $HW$. Throughout this section, $\Gamma$ has this conditional meaning. For any additional observation $\Oscr$, define
$
 \MMSE(\Oscr):=\E[\Var(V_T\mid c_{\rm pub},\Oscr)]$.

With one proposer and a known reserve, the observer sees only whether a sale occurs: whenever the proposer participates, its bid is the reserve. At location $x$, its post-public posterior-mean innovation has variance $I_H(x):=HW-D(x)$.

\begin{restatable}[Information-optimal monopoly]{proposition}{restatepropinfomonopoly}
\label{prop:info-monopoly}
For $I_H(x)>0$, the information-optimal reserve conditional on the public signal is $L=c_{\rm pub}$. It gives
\begin{equation}
 \MMSE_M^*(x)=D(x)+\alpha\,[HW-D(x)],
 \qquad \alpha:=1-\frac2\pi.
 \label{eq:monopoly-mmse}
\end{equation}
The information-optimal location minimizes $D$. Writing $D_*:=\min_x D(x)$, the optimal benchmark is $\alpha HW+(1-\alpha)D_*$.
\end{restatable}

\begin{proof}
For $I:=I_H(x)>0$, write the posterior mean as $c_{\rm pub}+\sqrt I\,Z$. The sale decision is $Z\geq z$, where $z=(L-c_{\rm pub})/\sqrt I$. Its explained posterior-mean variance is
\[
 I\frac{\phi(z)^2}{\Phi(z)\Phi(-z)}.
\]
This fraction is uniquely maximized at $z=0$, with value $2/\pi$; Appendix~\ref{app:info-monopoly} gives a direct proof. The independent residual variance $D(x)$ cannot be explained by the decision. Formula~\eqref{eq:monopoly-mmse} follows, as does the location claim. If $I=0$, the formula still holds and every reserve gives the same information.
\end{proof}

\subsection{The information in a winning bid}

With two proposers, consider the unrestricted no-reserve auction. Given the equilibrium characterization in \Cref{cor:gaussian} and \Cref{prop:degenerate}, the winning bid is a known function of the common information and the winner's standardized signal, so it reveals a combination of the two. The observer also learns that the loser's signal was at a lower quantile than the winner's.

Let $P$ and $J$ be the winning bid and winner identity.
Suppose first that $S:=s_A+s_B>0$. Use independent standard normals $Z_A,Z_B$ for the standardized signals, or for the virtual draw in Proposition~\ref{prop:degenerate} when a variance is zero. Let $M_2:=\max\{Z_A,Z_B\}$. Then
\begin{equation}
P=C-S\lambda(M_2).
\label{eq:price}
\end{equation}
Winner identity is independent of $M_2$ and $C$, and each proposer wins with probability $1/2$.

The loser has a normal signal truncated at $M_2$. Define its conditional variance and the difference between the winning signal and the loser's conditional mean by
\[
 \nu(m):=1-m\lambda(m)-\lambda(m)^2,\qquad
 r(m):=m+\lambda(m).
\]
For a standard normal $N$ independent of $M_2$, define
\begin{equation}
 \kappa:=\E[\nu(M_2)],\qquad
 \mathcal J(\eta):=
 \E\!\left[\Var\!\left(r(M_2)\mid-\lambda(M_2)+\sqrt\eta N\right)\right].
 \label{eq:universal-information}
\end{equation}
These quantities are independent of the locations and source weights. Normal integration gives
$\kappa\approx0.519462$. Moreover, Appendix~\ref{app:information} establishes the bounds
\begin{equation}
 \mathcal J(0)=0,\qquad
 \mathcal J(\eta)\text{ is nondecreasing},\qquad
 \mathcal J(\eta)<2-\kappa-\frac4\pi
 \quad(0\leq\eta<\infty).
 \label{eq:J-bounds}
\end{equation}

\begin{restatable}[Winning-bid informativeness]{theorem}{restatethmwinninginformation}
\label{thm:winning-information}
For every location profile with $s_A+s_B>0$,
\begin{equation}
 \MMSE_2(x_A,x_B)
 =\Omega_0+\frac{\Omega_A+\Omega_B}{2}
   \left[\kappa+\mathcal J\!\left(\frac{\Gamma}{(s_A+s_B)^2}\right)\right].
 \label{eq:two-mmse}
\end{equation}
If $s_A=s_B=0$, both bid $C$ and $\MMSE_2=\Omega_0$. In particular, at every profile
\begin{equation}
 \MMSE_2\leq\Omega_0+\alpha(\Omega_A+\Omega_B),
 \label{eq:mmse-bound}
\end{equation}
with strict inequality whenever $\Omega_A+\Omega_B>0$.
\end{restatable}

\begin{proof}
On an $A$-win, conditioning additionally on $M_2=m$ leaves mean $-\lambda(m)$ and variance $\nu(m)$ for $Z_B$. From \eqref{eq:price}, $C=P+S\lambda(m)$, so
\begin{align*}
 \E[V_T\mid P,J=A,M_2=m,c_{\rm pub}]&=P+s_A r(m),\\
 \Var(V_T\mid P,J=A,M_2=m,c_{\rm pub})&=\Omega_0+s_B^2\nu(m).
\end{align*}
The conditional variance formula, averaged over $P$, gives
$\Omega_0+s_B^2\kappa+s_A^2\mathcal J(\Gamma/S^2)$.
The analogous expression on a $B$-win exchanges $A$ and $B$. Averaging over the two equally likely winners proves \eqref{eq:two-mmse}. If both exclusive variances vanish, observing $P=C$ leaves only $\varepsilon$. Finally, \eqref{eq:J-bounds} implies
$[\kappa+\mathcal J(\eta)]/2<1-2/\pi=\alpha$, yielding \eqref{eq:mmse-bound}.
\end{proof}

\subsection{Competition versus the monopolist benchmark}

\begin{restatable}[Equilibrium information relative to monopoly]{theorem}{restatethminformationcomparison}
\label{thm:information-comparison}
At every pure two-proposer location equilibrium of the selected no-reserve game,
\begin{equation}
 \MMSE_2^{\rm eq}\leq \MMSE_M^*,
 \label{eq:competition-information}
\end{equation}
where the right side is the information-optimal monopolist benchmark in Proposition~\ref{prop:info-monopoly}. The inequality is strict whenever $HW>D_*$, that is, whenever some feasible location receives information newer than the public signal with positive variance.
\end{restatable}

\begin{proof}
Let $x_*$ minimize $D$. Coordinatewise optimality of $\Omega_0$ at an equilibrium gives
\[
 \Omega_0(x_A,x_B)\leq\Omega_0(x_*,x_B)\leq D_*.
\]
Using \eqref{eq:mmse-bound} and conditional variance accounting,
\begin{align*}
 \MMSE_2^{\rm eq}
 &\leq\Omega_0+\alpha(\Omega_A+\Omega_B)\\
 &=\alpha HW+(1-\alpha)\Omega_0-\alpha\Gamma\\
 &\leq\alpha HW+(1-\alpha)D_*=\MMSE_M^*.
\end{align*}
If there is any exclusive variance, the first inequality is strict. Otherwise $\MMSE_2^{\rm eq}=\Omega_0\leq D_*$, while the monopolist benchmark is strictly greater than $D_*$ when $HW>D_*$. This proves strictness in the remaining case.
\end{proof}

\subsection{Information-optimal locations}

Equilibrium locations need not be the most informative ones. We can characterize this difference exactly with two sources.

\label{app:info-locations}

Consider sources at $0$ and $1$, weights $w_0\geq w_1>0$, distance delays, and $H\geq1$. Write $m:=w_1$ and $W:=w_0+w_1$. For locations $0\leq a<b\leq1$, set $\Delta:=b-a$. Then
\begin{align}
 \Omega_A&=w_0\Delta,& \Omega_B&=w_1\Delta,\label{eq:two-source-info-vars}\\
 \Omega_0&=w_0a+w_1(1-b),&
 \Gamma&=w_0(H-b)+w_1(H-1+a).\nonumber
\end{align}
Define
\begin{equation}
 \eta_*:=\frac{W(H-1)}{(\sqrt{w_0}+\sqrt{w_1})^2},\qquad
 E_{\rm end}:=\frac W2[\kappa+\mathcal J(\eta_*)].
 \label{eq:endpoint-error}
\end{equation}
Co-location at the stronger source gives error $m$; opposite endpoints give error $E_{\rm end}$.

\begin{proposition}[Co-location or full separation]
\label{prop:information-locations}
The minimum two-proposer mean-square error over all locations is
$
 \min\{m,E_{\rm end}\}.$
Opposite endpoints are information-optimal exactly when $E_{\rm end}\leq m$. Co-location at the stronger source is information-optimal exactly when the reverse inequality holds; at equality both are optimal. For $H=1$, separation is optimal exactly when
\begin{equation}
 \frac{w_0}{w_1}\leq\frac2\kappa-1\approx2.850.
 \label{eq:information-threshold}
\end{equation}
For equal source weights, full separation is strictly better than co-location for every finite $H\geq1$.
\end{proposition}

\begin{proof}
Co-location at $a$ has error $D(a)\geq m$. For $a<b$,
\[
 \Omega_0=m(1-\Delta)+(w_0-w_1)a\geq m(1-\Delta).
\]
Also $\Gamma\geq W(H-1)$ and $(s_A+s_B)^2=\Delta(\sqrt{w_0}+\sqrt{w_1})^2$, so $\Gamma/(s_A+s_B)^2\geq\eta_*$. Monotonicity of $\mathcal J$ and Theorem~\ref{thm:winning-information} give
\[
 \MMSE_2(a,b)\geq (1-\Delta)m+\Delta E_{\rm end}
 \geq\min\{m,E_{\rm end}\}.
\]
Both benchmark errors are attained, proving the optimal value and the criterion. When $H=1$, $\eta_*=0$ and $\mathcal J(0)=0$, which gives \eqref{eq:information-threshold}. If $w_0=w_1=w$, then \eqref{eq:J-bounds} yields $E_{\rm end}<w(2-4/\pi)<w$.
\end{proof}

The location equilibrium always separates the proposers (\Cref{prop:two-sources}), so with
lopsided weights the geography that competition produces is not the most informative one.
Competition changes both what the proposers learn and how bidding discloses it.

\draftpagebreak

\bibliographystyle{plainnat}
\bibliography{references}

\appendix

\section{The regular monotone auction equilibrium}
\label{app:uniqueness}

This appendix proves the uniqueness statement in Remark~\ref{rem:selection}. Fix common information $c$, and suppose both bid functions are continuously differentiable and strictly increasing with positive derivatives on a common open bid support. Assume positive signal densities, so the quantiles of the inverse bid functions, denoted $q_A(b),q_B(b)$, are continuously differentiable with positive derivatives. These are regularity assumptions on the class in which uniqueness is asserted; they are not required for the existence proof of Theorem~\ref{thm:quantile}.

At a bid $b$ in the interior of the common support, a first-order condition for a type whose equilibrium bid is $b$ gives
\begin{align*}
 [g(q_A(b),q_B(b))-b]q_B'(b)&=q_B(b),\\
 [g(q_A(b),q_B(b))-b]q_A'(b)&=q_A(b).
\end{align*}
Both marginal types have the same common value at this signal pair. Dividing the equations yields
\[
 \frac{q_A'(b)}{q_A(b)}=\frac{q_B'(b)}{q_B(b)}.
\]
Thus the two inverse quantiles are proportional. At the common upper support endpoint they both tend to one, so $q_A(b)=q_B(b)$. Denote their common inverse by $q(b)$, and its inverse by $\beta(q)$. Either first-order condition becomes
\[
 q\beta'(q)=h(q)-\beta(q).
\]
Since $h$ is integrable, its solutions are
\[
 \beta(q)=\frac1q\int_0^q h(s)\,ds+\frac{k}{q}
\]
for a constant $k$. Relative to \eqref{eq:general-bid}, this subtracts $k$ from the interim profit of every quantile of either proposer. The base profit tends to zero at the lowest quantile, so zero limiting rent forces $k=0$.

For completeness, when $k<0$ the modified bid is still strictly increasing and the global deviation derivatives in Theorem~\ref{thm:quantile} are unchanged. It adds the same positive amount $-k$ to every interim profit. This illustrates why a lower-tail condition is needed with unrestricted negative bids. The construction is not ruled out by abstention alone. A positive $k$ cannot give an increasing finite-valued bid on the full quantile interval, since $q\beta_0(q)\to0$ for the base solution while $k/q$ diverges upward as $q\downarrow0$.

\begin{example}[A symmetric product]
\label{ex:product}
Let $X_A,X_B$ be independent and uniform on $0<x<1$, and let $V=X_AX_B$. Then $h(q)=q^2$, so $\beta(q)=q^2/3$. Each proposer's interim rent is $q^3/6$, its ex ante rent is $1/24$, and expected revenue is $1/6$. In particular, $\beta(q)$ is not the value conditional on two independently truncated signals: that value is $q^2/4$. Formula~\eqref{eq:general-bid} averages along \emph{matched quantiles}, not over the square of independently truncated quantiles.
\end{example}

\section{Reserve-auction proofs and boundary cases}
\label{app:reserve}

\subsection{Proof of Theorem~\ref{thm:reserve-general}}
\label{app:reserve-proof}

Use the $H,W$ ordering and the notation of Section~\ref{sec:reserve}. Strict monotonicity gives
\[
 h(q_R)=g(q_R,q_R)>
 \frac1{q_R}\int_0^{q_R}g(s,q_R)\,ds=L.
\]
For $q>q_R$, differentiating \eqref{eq:reserve-general-bid} gives
\[
 q^2\beta_L'(q)
 =q_R[h(q)-L]+\int_{q_R}^q[h(q)-h(s)]\,ds>0.
\]
Thus the bid branch is strictly increasing, starts at $L$, and has a finite upper endpoint by integrability of $h$.

\emph{Deviations on the increasing branch.}
A bid associated with $r>q_R$ defeats exactly the opponent quantiles below $r$, including all its inactive or pooled types. A type $q$ of proposer $H$ therefore receives
\[
 G_H(r;q)=\int_0^r g(q,s)\,ds-r\beta_L(r),
 \qquad \partial_rG_H(r;q)=g(q,r)-g(r,r).
\]
The derivative has the sign of $q-r$. For $W$, the corresponding derivative is $g(r,q)-g(r,r)$ and has the same sign. Consequently, types above $q_R$ prefer their own branch bid to every other branch bid. Types below $q_R$ prefer the limiting boundary bid to any strictly higher bid.

\emph{Entry and pooling by $H$.}
Proposer $W$ has no atom at $L$. A bid of exactly $L$ by $H$ therefore wins whenever $W$'s quantile is below $q_R$, yielding
\[
 U_H(L;q)=\int_0^{q_R}[g(q,s)-L]\,ds.
\]
This is strictly increasing in $q$. By the definition of $p_R$, it is negative below $p_R$ and nonnegative above it, with the evident endpoint convention if $p_R=0$. Its value is also the limiting profit on the increasing branch as $r\downarrow q_R$. The branch deviation comparison therefore gives precisely abstention below $p_R$, pooling from $p_R$ to $q_R$, and the increasing branch above $q_R$.

\emph{Entry and deviations to $L$ by $W$.}
At $W$'s cutoff type $q_R$, a bid just above $L$ yields zero by \eqref{eq:general-entry}. Lower types obtain a strictly negative limiting profit and cannot gain on the increasing branch. To check a bid of exactly $L$, put
\[
 f(s):=g(s,q_R)-L,\qquad
 \int_0^{q_R}f(s)\,ds=0.
\]
The function $f$ is increasing. It follows that $\int_0^{p_R}f(s)\,ds\leq0$: any lower prefix has average at most that of the entire interval. If a reserve tie is won with probability $\theta$ (equal to $1/2$ under our rule), the cutoff type's profit from bidding $L$ is
\[
 (1-\theta)\int_0^{p_R}f(s)\,ds
       +\theta\int_0^{q_R}f(s)\,ds\leq0.
\]
Lower $W$ types obtain no more, so abstention is optimal for them.

For a type $q\geq q_R$, the difference between the limiting profit from just above $L$ and the profit from exactly $L$ is
\[
 (1-\theta)\int_{p_R}^{q_R}[g(s,q)-L]\,ds.
\]
At $q=q_R$ this is nonnegative because it is minus $1-\theta$ times the nonpositive prefix integral above. It is increasing in $q$. Thus an active $W$ type does not prefer a deviation to the reserve; its own branch bid already weakly dominates the limiting boundary bid.

Abstention gives zero. Active types have nonnegative profit by the entry comparisons. Bids above the upper branch endpoint only pay more for a certain win. These observations cover all remaining deviations and complete the equilibrium proof. No sale requires $H$'s quantile below $p_R$ and $W$'s below $q_R$, giving probability $p_Rq_R$. A sale at $L$ requires $H$ in its pool and $W$ inactive, giving $(q_R-p_R)q_R$.

\subsection{Normal integrals and the zero-variance cases}
\label{app:reserve-degenerate}

The normal integrals used in Proposition~\ref{prop:reserve-profits} can be evaluated without a further expectation over private signals:
\begin{align*}
 A(z)&=\frac{\Phi(-\sqrt2 z)}{\sqrt\pi}
       -z\Phi(z)\Phi(-z)-\phi(z)[1-2\Phi(z)],\\
 B(u,z)&=z\Phi(-z)-\phi(z)-u\Phi(-u)+\phi(u).
\end{align*}
The second identity follows by differentiation. For the first, integrate $\Phi(t)\Phi(-t)$ by parts and use $\phi'(t)=-t\phi(t)$ and the normal integral of $\phi(t)^2$. In particular, $A(-\infty)=1/\sqrt\pi$.

When $s_W=0$, the continuations below give the Gaussian counterpart of the reserve equilibrium of Moallemi, Pai, and Robinson~\cite{moallemi2026latency}, with the uninformed bidder abstaining rather than pooling at the reserve to accommodate our tie-breaking rule. This is distinct from $p_R=0$, which means that every informed type enters, not that the other bidder is uninformed.

Suppose now that $s_H=s>0$ and $s_W=0$, conditional on $C=c$. The following strategies define reserve continuations at these degenerate profiles.

\emph{If $c\leq L$,} proposer $W$ abstains. Proposer $H$ bids $L$ if $c+X_H\geq L$ and otherwise abstains. This is an equilibrium. An uninformed bid strictly above $L$ wins for sure and has expected value $c$; a bid of exactly $L$ gives no larger expected profit, since it is relatively more likely to win when $H$'s signal is unfavorable. The conditional profits are
\[
 \pi_H(c)=(c-L)\Phi((c-L)/s)+s\phi((c-L)/s),
 \qquad \pi_W(c)=0.
\]

\emph{If $c>L$,} let $z_R$ solve $\lambda(z_R)=(c-L)/s$, put $q_R=\Phi(z_R)$, and let $u:=-\lambda(z_R)<z_R$. Proposer $H$ abstains below standardized signal $u$, bids $L$ between $u$ and $z_R$, and bids $c-s\lambda(z)$ above $z_R$. Proposer $W$ draws a uniform $Q$, abstains if $Q\leq q_R$, and otherwise bids $c-s\lambda(\Phi^{-1}(Q))$.

For every bid on the increasing branch, the uninformed proposer's conditional expected value on winning equals its bid, so its profit is zero. A bid of exactly $L$ has nonpositive profit by the same prefix-integral argument as above. The informed proposer's opponent distribution has cumulative probability $q$ at branch quantile $q$, so its global branch comparison and reserve-pooling comparison remain valid. The conditional profits are
\[
 \pi_H(c)=s[A(z_R)+q_R B(u,z_R)],\qquad \pi_W(c)=0.
\]
These are actual mixed equilibria, not just limits of formulas with positive $s_W$.

Finally, if both exclusive variances vanish, both proposers abstain when $c<L$ and both bid $c$ when $c\geq L$. Conditional profit is zero for each proposer. Together these cases define the payoff in \eqref{eq:reserve-location} at every Brownian location profile.

\section{Choosing the auctioneer's location}
\label{sec:auctioneer-choice}
\label{app:more-location}

Fix proposer locations and let an auctioneer at $z$ create delivery delays $\rho(x_A,z)$ and $\rho(x_B,z)$. Write
\[
 \Delta_k:=\delta_k(x_B)-\delta_k(x_A),\qquad
 h_a(z):=\rho(x_B,z)-\rho(x_A,z).
\]
Under the untruncated-delay assumption, the exclusive variances are
\begin{equation}
 \Omega_A(h)=\sum_k w_k[\Delta_k+h]_+,\qquad
 \Omega_B(h)=\sum_k w_k[-\Delta_k-h]_+.
 \label{eq:auctioneer-variances}
\end{equation}
Thus location affects no-reserve revenue only through the scalar $h=h_a(z)$:
\begin{equation}
 R(h)=V_0-\frac{\sqrt{\Omega_A(h)}+\sqrt{\Omega_B(h)}}{\sqrt\pi}.
 \label{eq:auctioneer-revenue}
\end{equation}

\begin{proposition}[Revenue with fixed proposer locations]
\label{prop:auctioneer-revenue}
If the feasible differential latencies form a compact interval, a revenue-maximizing differential latency can be found among its endpoints and the feasible breakpoints $-\Delta_k$.
\end{proposition}

\begin{proof}
Between consecutive breakpoints, both variances in \eqref{eq:auctioneer-variances} are nonnegative affine functions of $h$. Their square roots are concave, so revenue is convex on that interval and attains a maximum at an endpoint. Comparing all such endpoints proves the claim.
\end{proof}

If proposers choose their locations after the auctioneer commits, this fixed-location calculation is a different game. Proposition~\ref{prop:auctioneer} supplies a continuation $(z,z)$ with revenue $V_0$, but also supplies continuations $(z,y)$ with revenue
\[
 V_0-\sqrt{D(y)-D(z)}/\sqrt\pi.
\]
Claiming that the auctioneer's commitment \emph{guarantees} revenue $V_0$ would therefore require an equilibrium-selection argument.

\section{A co-located equilibrium with one source and many proposers}
\label{app:one-source-many}

Let $M\geq3$, take one source at $0$, and let the feasible location set be $\{0,1\}$ with distance delays and $T\geq1$. Write $C:=V_0+\sigma B_{T-1}$ and $X:=\sigma(B_T-B_{T-1})$, so $V_T=C+X$. A proposer at $0$ observes $V_T$, while a proposer at $1$ observes $C$. Conditional on $C$, $X$ is independent normal with standard deviation $s=\sigma$.

We specify auction equilibria at every location profile. If at least two proposers are at $0$, they bid $V_T$ and all proposers at $1$ abstain. The informed proposers cannot earn a positive profit by deviating against another bid of $V_T$. A proposer at $1$ can only win with a bid at least $V_T$ and therefore cannot earn a positive profit either. If no proposer is at $0$, all bid $C$, an equilibrium among identically informed bidders.

If exactly one proposer is at $0$, let $n=M-1$ be the number at $1$. The informed proposer uses
\[
 \beta(q)=C-s\lambda(\Phi^{-1}(q)),\qquad q=\Phi(X/s).
\]
Each of the $n$ uninformed proposers independently draws a random quantile $Q$ with distribution function $\PR(Q\leq q)=q^{1/n}$ and bids $\beta(Q)$. The maximum of their draws is uniform, so the informed proposer faces exactly the opponent bid distribution of Proposition~\ref{prop:degenerate} and is best responding. At any support bid of an uninformed proposer, conditioning on defeating the informed proposer makes expected value equal to the bid; defeating the other uninformed proposers supplies no further information about $X$. Thus every uninformed support bid has zero expected profit, and bids at or above $C$ cannot improve on it. This is an auction equilibrium as well.

With these continuations, the location profile in which all proposers choose $0$ is a Nash equilibrium. Everyone receives zero profit, and a unilateral move to $1$ leaves at least two informed rivals at $0$ and still gives the deviator zero. This example shows why a no-co-location statement needs assumptions on the sources and on the available information advantage, even before considering the additional difficulties of general multi-proposer information structures.

\section{Details for price informativeness}
\label{app:information}

\subsection{The optimal binary threshold}
\label{app:info-monopoly}

For a standard normal $Z$, observing whether $Z\geq z$ explains variance
\[
 \Var(\E[Z\mid\I{Z\geq z}])
 =\frac{\phi(z)^2}{\Phi(z)\Phi(-z)}.
\]
Indeed, the conditional means are $\phi(z)/\Phi(-z)$ above the threshold and $-\phi(z)/\Phi(z)$ below it. To maximize the explained variance, use
\begin{align*}
 \frac{\Phi(z)\Phi(-z)}{\phi(z)^2}
 &=\int_0^\infty\int_0^\infty
   e^{-(u^2+v^2)/2} e^{z(u-v)}\,du\,dv\\
 &=\int_0^\infty\int_0^\infty
   e^{-(u^2+v^2)/2}\cosh(z(u-v))\,du\,dv\\
 &\geq\left(\int_0^\infty e^{-u^2/2}\,du\right)^2
 =\frac\pi2.
\end{align*}
The second equality averages the first integral with the one obtained by interchanging $u$ and $v$. The inequality is strict for $z\ne0$, since $\cosh(z(u-v))>1$ off the diagonal. This proves the unique optimal threshold in Proposition~\ref{prop:info-monopoly}.

\subsection{The universal normal quantities}

Let $M_2$ be the maximum of two independent standard normals. Its density is $f_M(m)=2\phi(m)\Phi(m)$. A normal truncated above at $m$ has mean $-\lambda(m)$ and variance $\nu(m)$, as follows from integrating $z\phi(z)$ and $z^2\phi(z)$ up to $m$. In particular, $\nu(m)>0$. Also,
\[
 \lambda'(m)=-\lambda(m)(m+\lambda(m))<0,
 \qquad r'(m)=\nu(m)>0.
\]
Thus $-\lambda$ and $r$ are both strictly increasing.

The identities required for \eqref{eq:J-bounds} are
\begin{equation}
 \E[M_2]=\E[\lambda(M_2)]=\frac1{\sqrt\pi},\qquad
 \E[M_2^2]=1,\qquad \E[M_2\lambda(M_2)]=0.
 \label{eq:normal-moments}
\end{equation}
For the first identity, integration by parts gives
$\E[M_2]=2\int\phi(m)^2\,dm=1/\sqrt\pi$, and cancellation of $\Phi$ gives the same integral for $\E[\lambda(M_2)]$. The squared maximum and squared minimum have the same distribution and sum to the sum of the two squared normals, proving the second moment. Finally, the mixed moment is $2\int m\phi(m)^2\,dm=0$.

From the definition of $\nu$, \eqref{eq:normal-moments} implies
$\E[\lambda(M_2)^2]=1-\kappa$. Hence
\begin{equation}
 \E[r(M_2)]=\frac2{\sqrt\pi},\qquad
 \E[r(M_2)^2]=2-\kappa,\qquad
 \Var(r(M_2))=2-\kappa-\frac4\pi.
 \label{eq:r-moments}
\end{equation}
The constant $\kappa$ is given by the absolutely convergent integral
\[
 \kappa=\int_{-\infty}^\infty
 [1-m\lambda(m)-\lambda(m)^2]2\phi(m)\Phi(m)\,dm
 \approx0.519462.
\]

For $\eta=0$, $-\lambda(M_2)$ determines $M_2$, so $\mathcal J(0)=0$. If $0\leq\eta_1<\eta_2$, the observation at $\eta_2$ can be obtained from the one at $\eta_1$ by adding an independent normal of variance $\eta_2-\eta_1$. Conditional expectation and the conditional variance identity then give $\mathcal J(\eta_1)\leq\mathcal J(\eta_2)$.

Total variance gives the weak bound $\mathcal J(\eta)\leq\Var(r(M_2))$. It is strict at every finite $\eta$. To see this, set $Y=-\lambda(M_2)+\sqrt\eta N$. Since $r$ and $-\lambda$ are strictly increasing and $M_2$ is nondegenerate,
\[
 \Cov(r(M_2),Y)=\Cov(r(M_2),-\lambda(M_2))>0.
\]
One can verify strict positivity by taking an independent copy and writing covariance as half the expectation of the product of the two corresponding differences. If $\E[r(M_2)\mid Y]$ were constant, this covariance would be zero. Therefore $\Var(\E[r(M_2)\mid Y])>0$, which proves the strict bound in \eqref{eq:J-bounds}.

\subsection{An integral representation for computation}

For $\eta>0$, define
\[
 K_\eta(y,m):=2\phi(m)\Phi(m)
       \frac1{\sqrt\eta}\phi\!\left(\frac{y+\lambda(m)}{\sqrt\eta}\right).
\]
This is the joint density of $Y=-\lambda(M_2)+\sqrt\eta N$ and $M_2$. Consequently,
\begin{equation}
 \mathcal J(\eta)=2-\kappa-
 \int_{-\infty}^\infty
 \frac{\left(\int_{-\infty}^\infty r(m)K_\eta(y,m)\,dm\right)^2}
      {\int_{-\infty}^\infty K_\eta(y,m)\,dm}\,dy.
 \label{eq:J-integral}
\end{equation}
The denominator is positive for every finite $y$. This formula, together with $\mathcal J(0)=0$, evaluates the information criterion without simulating bidding strategies.

\end{document}